\documentclass[preprint,showpacs,showkeys,preprintnumbers,amsmath,amssymb,superscriptaddress]{revtex4}

\usepackage{graphicx}
\usepackage{dcolumn}
\usepackage{bm}
\usepackage[latin1]{inputenc}

\begin{document}

\author{R. Rossi Jr.}
\affiliation{Departamento de F\'{\i}sica, Instituto de Ciências
Exatas, Universidade Federal de Minas Gerais, C.P. 702, 30161-970,
Belo Horizonte, MG, Brazil}

\author{A. R. Bosco de Magalh\~{a}es}
\affiliation{Departamento de F\'{\i}sica e Matem\'{a}tica, Centro
Federal de Educação Tecnológica de Minas Gerais, Av. Amazonas, 7675,
30510-000, Belo Horizonte, MG, Brazil}

\author{M. C. Nemes}

\affiliation{Departamento de F\'{\i}sica, Instituto de Ciências
Exatas, Universidade Federal de Minas Gerais, C.P. 702, 30161-970,
Belo Horizonte, MG, Brazil}

\title{Incomplete Entanglement: consequences for the QZE}

\begin{abstract}
We show that the Quantum Zeno Effect prevails even if the
entanglement with the measuring probe is not complete. The dynamics
towards the asymptotic regime as a function of $N$, the number of
measurements, reveals surprising results: the transition
probability, for some values of the coupling to the measuring probe,
may decrease much faster than the normal QZE.
\end{abstract}

\maketitle

The development of Quantum Information Theory caused a surprisingly
great enhancement in the research of the Quantum Zeno Effect. In
fact the citations of the famous work by B. Misra and E. C.
Sudarshan \cite{art1}, and Itano et. al. \cite{art2} has
impressively increased as shown in \cite{art3}. The reason for this
is connected to the possibility realized by the QZE of controlling
and preserving quantum states \cite{art4,art5,art6,art7,art8}.

In this contribution we investigate modifications on the dynamics of
a quantum system provoked by interactions with probe systems. The
system we use is that of two linearly coupled qubits one of which is
coupled to the measuring apparatus at periodic times. We show that
the QZE persists even if the system-measuring apparatus (probe) do
not entangle completely after each interaction. This means that the
state of the measuring apparatus after each interaction does not
contain conclusive information about the system's state. Physically
this means that the relative importance of each correlation process
becomes less and less important as $N$ increases. Moreover we show
that it is possible to control the subsystem (two qubits) through
interactions which do not entangle (at all) with the auxiliary
probes. However, surprisingly enough the initial state is more
efficiently protected than by the traditional QZE.

\emph{The system}

Let us consider a system $S$ composed by two coupled qubits ($S_{a}$
and $S_{b}$) described by the Hamiltonian:

\begin{equation}
H_{S}= \epsilon_{a}|1_{a}\rangle\langle 1_{a}| +
\epsilon_{b}|1_{b}\rangle\langle 1_{b}| +\hbar
G(\sigma^{a}_{-}\sigma^{b}_{+} + \sigma^{a}_{+}\sigma^{b}_{-}),
\end{equation}
where $\sigma_{+}=|1\rangle\langle 0|$, $\sigma_{-}=|0\rangle\langle
1|$, $G$ is the coupling coefficient and $\epsilon_{a}$
($\epsilon_{b}$) is the eigenvalue of $S_{a}$ ($S_{b}$) free
Hamiltonian. A possible empirical implementation for this model can
be performed in a solid-state superconducting device \cite{art9}.

Let us consider a measuring system consisting of a set of two level
systems (whose states are represented by $|1^{(k)}_{M}\rangle$ and
$|0^{(k)}_{M}\rangle$) that interact with $S_{b}$. This interaction
will discriminate between the states $|1_{b}\rangle$ and
$|0_{b}\rangle$. For simplicity we will follow the analysis
considering instantaneous interactions, described by the
hamiltonian:

\begin{eqnarray}
H^{(k)}_{SM}(t)&=&I_{a}\otimes g \hbar
\delta(t-t^{(k)}_{m})(\sigma^{b}_{+}|0^{(k)}_{M}\rangle\langle
1^{(k)}_{M}|
+\sigma^{b}_{-}|1^{(k)}_{M}\rangle\langle 0^{(k)}_{M}|),\\
&=&I_{a}\otimes g \hbar \delta(t-t^{(k)}_{m})\Gamma^{(k)}.
\end{eqnarray}

The delta function limits the interaction times to $t^{(k)}_{m}$,
$g$ is the interaction coefficient and $I_{a}$ represents the
identity matrix of subsistem $S_{a}$. In what follows we conclude
that the coefficient $g$ is related to the completeness of the
measurement. The total Hamiltonian ($S+M$) can be written as:

\begin{equation}
H(t)=H_{S}+H_{M}+\sum^{N}_{k=1}H^{(k)}_{SM}(t),
\end{equation}
where $H_{M}$ is the free Hamiltonian of the probe system

\begin{equation}
H_{M}=\sum_{k=1}^{N}\left(\lambda^{(k)}_{1}|1^{(k)}_{M}\rangle\langle
1^{(k)}_{M}|+\lambda^{(k)}_{0}|0^{(k)}_{M}\rangle\langle
0^{(k)}_{M}|\right),
\end{equation}
$\lambda^{(k)}_{1}$($\lambda^{(k)}_{0}$) is the energy of the
eigenvector $|1^{(k)}_{M}\rangle$($|0^{(k)}_{M}\rangle$). Let us
divide the time evolution of the initial state in $N$ steps, and
each step in two parts. At the first one, $S_{a}$ and $S_{b}$
interact freely, this evolution being governed by:

\begin{equation}
U_{S}(t)=\exp\left(\frac{-i}{\hbar}(H_{S}+H_{M})t\right).\label{evo1}\\
\end{equation}
This first part is identical in every step. At the second part, the
probe system interacts with $M_{b}$, and the unitary evolution on
the $k^{th}$ step is given by

\begin{eqnarray}
U^{(k)}_{SM}(t^{(k)}_{m}-\epsilon,
t^{(k)}_{m}+\epsilon)&=&\exp\left(\frac{-i}{\hbar}\int_{t^{(k)}_{m}-\epsilon}^{t^{(k)}_{m}+
\epsilon}(H_{S}+H_{M}+\sum_{k=1}^{N}H^{(k)}_{SM}(t))dt\right),\\
&=&\exp\left(\frac{-i}{\hbar}(H_{S}+H_{M})2\epsilon-i g
\Gamma^{(k)}\right),
\end{eqnarray}
taking the limit $\epsilon \rightarrow 0$, we have
$U^{(k)}_{SM}(t^{(k)}_{m}-\epsilon,
t^{(k)}_{m}+\epsilon)=\exp\left(-i g\Gamma^{(k)}\right)$. Note that
$\Gamma^{(k)}$ has the following properties:
$(\Gamma^{(k)})^{n}=\Gamma^{(k)}$ when $n$ is an odd number, and
$(\Gamma^{(k)})^{n}=I$ when $n$ is even. Therefore, from the series
expansion of $U^{(k)}_{SM}(t^{(k)}_{m}-\epsilon,
t^{(k)}_{m}+\epsilon)$ we may write

\begin{equation}
U^{(k)}_{SM}(t^{(k)}_{m}-\epsilon, t^{(k)}_{m}+\epsilon)
=\exp\left(-i g\Gamma^{(k)}\right) =I\cos(g)-i\sin(g)\Gamma^{(k)}.
\label{evo2}
\end{equation}
The total time evolution of the global system will be composed by a
succession of the unitary evolutions shown in (\ref{evo1}) and
(\ref{evo2}) as follows:

\begin{equation}
|\psi(T)\rangle=(U^{(N)}_{SM}U_{S})...(U^{(k)}_{SM}U_{S})(U^{(k-1)}_{SM}U_{S})...(U^{(1)}_{SM}U_{S})|\psi(0)\rangle.
\end{equation}

As will become clear in what follows our analysis is along the lines
of \cite{art10}.

\emph{Single interaction}

Let us consider $|\psi(0)\rangle=|1,0\rangle|0_{M}\rangle$ as the
initial state of the system ($S+M$), where
$|0_{M}\rangle=\bigotimes_{k=1}^{N}|0^{k}_{M}\rangle$. At the first
part of the first step it evolves as:

\begin{equation}
|\psi(t^{(1)}_{m})\rangle=U_{S}(t^{(1)}_{m})|\psi(0)\rangle=\left(\alpha(t^{(1)}_{m})|1,0\rangle-i\beta(t^{(1)}_{m})|0,1\rangle\right)|0_{M}\rangle,
\end{equation}
where $\alpha(t)=\cos(Gt)$ and $\beta(t)=\sin(Gt)$. Note that the
quantum transition that will be modified by the interactions with
the probe is: $|1,0\rangle\rightarrow |0,1\rangle$.

At the second part of the first step, subsystem $S_{b}$ interacts
with the first probe (we omitted states of the probe that do not
interact in this step):

\begin{equation}
U^{(1)}_{SM}|\psi(t^{(1)}_{m})\rangle=\left(\alpha(t^{(1)}_{m})|1,0\rangle-i\beta(t^{(1)}_{m})\cos(g)|0,1\rangle\right)|0^{(1)}_{M}\rangle
-\beta(t^{(1)}_{m})\sin(g)|0,0\rangle|1^{(1)}_{M}\rangle.
\label{eq3}
\end{equation}

Information about occurrence of the quantum transition
($|1,0\rangle\rightarrow |0,1\rangle$) may be obtained through the
probe state. If $g=\frac{k\pi}{2}$ (where $k=1,2,3...$), information
is complete. The probe state is $|1^{(1)}_{M}\rangle$ if the
transition took place and $|0^{(1)}_{M}\rangle$ if it did not.
However, if $g=\pi$, the system does not entangle with the probe,
therefore it does not carry any information about the quantum
transition occurrence. For other values of $g$ we have intermediate
configurations (incomplete information). It is clear that
$|\cos(g)|$ quantifies the amount of information.

\emph{The transition rate}

In this section we investigate the changes on the quantum transition
rate roused by the interaction described in (\ref{eq3}).

An approach for the Quantum Zeno effect that focus on the changes of
the quantum transition rate is presented in \cite{art11}. The
authors show that immediately after a complete measurement
(interaction that induces the maximum entanglement between $S$ and
$M$), the quantum transition rate is null. Therefore, a complete
measurement series inhibits the enhancement of $\frac{dP_{in}}{dt}$,
and consequently, of the quantum transition. We investigate, in this
section, the effects of incomplete measurements and interactions
that do not entangle $S$ and $M$ at all.

Firstly, let us calculate the quantum transition rate for an
interval where no measurements are performed.

\begin{eqnarray}
T^{(0)}(t)&=& \frac{dP_{1,0}}{dt}=\frac{d}{dt}\left[\langle
\psi(0)|U_{S}^{\dagger}(t)(|1,0\rangle\langle 1,0|\otimes I_{M})|U_{S}(t)|\psi(0)\rangle\right]\\
&=&-2G\alpha(t)\beta(t),
\end{eqnarray}
notice that the rate is null when $t=0$  and it assumes non null
values for later times, allowing the states in subsystem $S$ to
evolve. Let us calculate the rate immediately after an interaction
with $M$. For this purpose, we consider the vector state at
$t^{(1)}_{m}+t$, where $t^{(1)}_{m}$ (as defined previously) is the
instant of the first  interaction with $M$ and $t$ a time interval
after this interaction,where the system evolves freely.

\begin{eqnarray}
|\psi(t^{(1)}_{m}+t)\rangle&=&\left(U_{S}(t)U^{(1)}_{SM}U_{S}(t^{(1)}_{m}))\right)|1,0\rangle|0_{M}\rangle \notag\\
&=&\left(\alpha(t^{(1)}_{m})\alpha(t)-\beta(t^{(1)}_{m})\beta(t)\cos(g)\right)|1,0\rangle|0^{(1)}_{M}\rangle
\notag \\
&&-i\left(\alpha(t^{(1)}_{m})\beta(t)-\beta(t^{(1)}_{m})\alpha(t)\cos(g)\right)|0,1\rangle|0^{(1)}_{M}\rangle\notag\\
&&-\beta(t^{(1)}_{m})\sin(g)|0,0\rangle|1^{(1)}_{M}\rangle.
\end{eqnarray}

We calculate the quantum transition rate as function of $t$ and take
the limit $t\rightarrow 0$:

\begin{eqnarray}
T^{(1)}(t^{(1)}_{m})&=&\lim_{t\rightarrow
0}\frac{dP_{1,0}}{dt}=\frac{d}{dt}\left(\langle
\psi(t^{(1)}_{m}+t)|(|1,0\rangle\langle 1,0|\otimes I_{M})|\psi(t^{(1)}_{m}+t)\rangle\right)\\
&=&-2G\alpha(t^{(1)}_{m})\beta(t^{(1)}_{m})\cos(g),\label{eq4}
\end{eqnarray}
then we get the quantum transition rate immediately after the
interaction between $S$ and $M$. Different modifications on the
quantum transition rate can be observed for different values of $g$.
As $\cos(g)$ is a periodic function, let us restrict ourselves to
the interval $0\leq g\leq \pi$. When $g=0$ there is no interaction,
we can notice that $T^{(0)}(t)=T^{(1)}(t)$.

For $0< g<\frac{\pi}{2}$ there is a decrease of the quantum
transition rate's absolute value
$|T^{(0)}(t^{(1)}_{m})|>|T^{(1)}(t^{(1)}_{m})|$, however this
interaction with $M$ does not invert the ``signal" of the derivative
(quantum transition rate). A similar work along these lines that
focus on this particular interval is \cite{art12}.

If $g=\frac{\pi}{2}$ (complete measurement) the transition rate is
null after the interaction with  $M$, $T^{(1)}(t^{(1)}_{m})=0$
(traditional QZE).

When $\frac{\pi}{2}< g<\pi$, we can notice a decrease of the quantum
transition  rate's absolute value
$|T^{(0)}(t^{(1)}_{m})|>|T^{(1)}(t^{(1)}_{m})|$ and also a
derivative's ``signal" inversion. We know that the derivative signal
decides whether the function $P_{1,0}(t)$ is increasing or
decreasing. Therefore, after this interaction $P_{1,0}(t)$ becomes
increasing. It is interesting to notice that this interactions does
not entangle $S$ and $M$ completely, nevertheless, it inhibits the
transition all the same in a stronger sense than the traditional
QZE.

For $g=\pi$, only the inversion of the derivative's signal is
observed, there is no change on its absolute value. This interaction
has a net result similar to the inverting pulse in the ``Super-Zeno
Effect"\cite{art13}.

In Fig 1., we see the probability curves $P^{(1)}_{1,0}$ as a
function of time for different values of $g$. The time interval
considered is divided by one interaction with $M$, that happens at
$t^{(1)}_{m}=0,5$. The thick continuous curve represents
$P^{(1)}_{1,0}$ without probe intervention, notice that all the
other curves move away from this one, exhibiting the fact that any
interaction ($g\neq 0$) inhibits the quantum transition.

\begin{figure}[h]
\centering
%\hspace{-3.5cm}
  \includegraphics[scale=0.5]{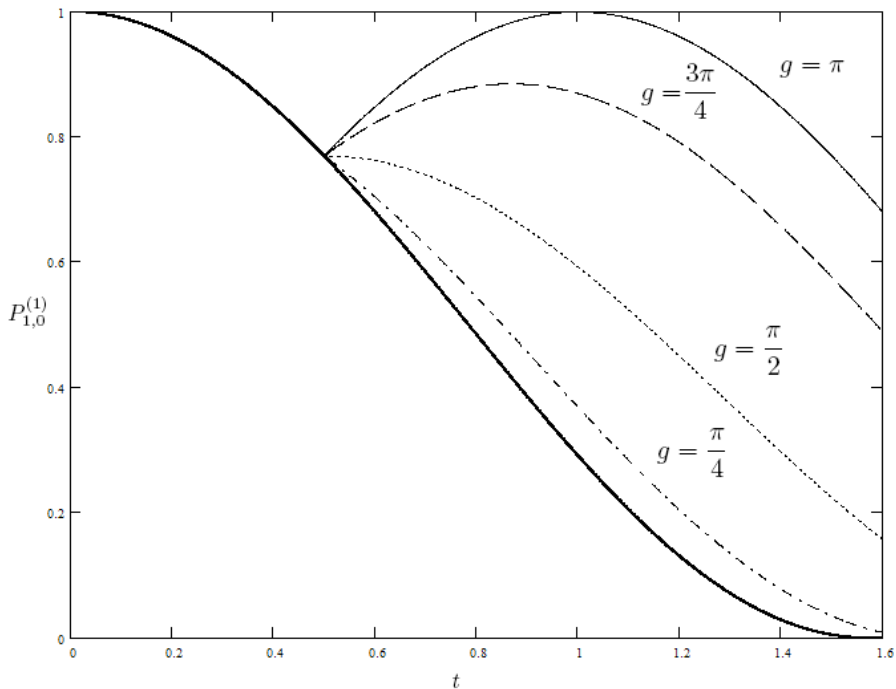}\\
  \caption{$P^{(1)}_{1,0}\times t$ for different values of  $g$.}
\end{figure}

The strongest inhibition is roused by $g=\pi$ ($\pi$ pulse), which
inverts the derivative signal without changing its absolute value.
After some algebra we may write the quantum transition rate as:

\begin{equation}
T^{(1)}(t^{(1)}_{m}+t,g=\pi)=\sin(2t^{(1)}_{m}-2t).
\end{equation}

The rate is positive ($T^{(1)}(t^{(1)}_{m}+t,g=\pi)>0$) when the
evolution time after the interaction with $M$ is smaller than the
evolution time before the interaction ($t^{(1)}_{m}>t$). Therefore,
the probability $P^{(1)}_{1,0}$ is an increasing function in this
interval. For $\frac{\pi}{2}<g<\pi$ the time interval when
$P_{1,0}(t)$ is increasing is smaller, that is the reason for the
strongest inhibition roused by $g=\pi$.

\emph{$N$ sequential measurements}

Let us focus on the investigation of a sequence of $N$ interactions
between $S$ and $M$ for different values of $g$.

Firstly, we consider interactions that do not entangle the
subsystems $S$ and $M$, but rouses meaningful changes on the
evolution of $S$ ($g=\pi$).

After, we investigate the possibility of Quantum Zeno Effect with
incomplete measurements. We conclude that the enhancement on the
number of interactions ($N$) contribute to the decreasing of the
quantum transition rate's absolute value. This is the physical
mechanism which allows for the similar behavior between incomplete
and complete measurements when $N\rightarrow\infty$.

The dynamics of the system $S$ may be controlled through a sequence
of interactions with $g=\pi$.

The modifications induced by a sequence of $N$ $g=\pi$ interactions
in $T=N\tau$, where $\tau$ is the time period during which $S$
evolves freely, depends on the parity of $N$. Because at the end of
each $g\pi$ interaction the signal of $\frac{dP_{1,0}}{dt}$ is
inverted, alternating the behavior of $P_{1,0}(t)$ between
increasing and decreasing.

If $N$ is even

\begin{equation}
\prod_{k=1}^{N}\left(U^{k}_{SM}U_{S}\right)|1,0\rangle|0_{M}\rangle=|1,0\rangle|0_{M}\rangle,\label{eq5}
\end{equation}
and if $N$ is odd
\begin{equation}
\prod_{k=1}^{N}\left(U^{k}_{M}U_{S}\right)|1,0\rangle|0_{M}\rangle=\left(\alpha(\tau)|1,0\rangle+\beta(\tau)|0,1\rangle\right)|0_{M}\rangle.\label{eq6}
\end{equation}

The control of the dynamics can be achieved from equations
(\ref{eq5}) and (\ref{eq6}). Notice that quantum transition
inhibition is more efficient for a sequence of $g=\pi$ interactions
than for the traditional QZE.

\begin{figure}[h]
\centering
%\hspace{-3.5cm}
  \includegraphics[scale=0.65]{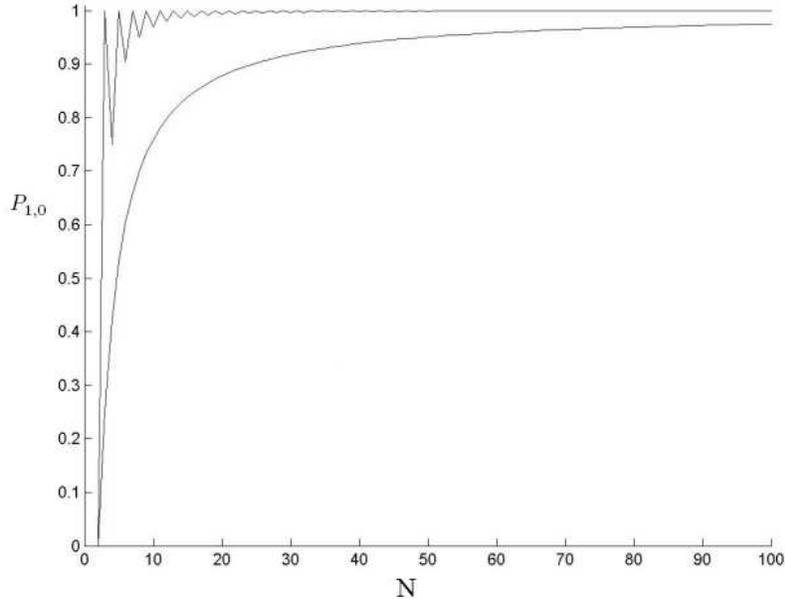}\\
  \caption{$P_{1,0}\times N$ for $g=\pi$ (upper curve), $g=\frac{\pi}{2}$(lower curve) and $T=\frac{\pi}{2g}$.}
\end{figure}

The presence of oscillations in the upper curve is due to the
alternate behavior of $P_{1,0}(t)$ (increasing/decreasing) after
each $S-M$ interaction. When $N\rightarrow\infty$ the oscillation
disappears since the state vector (\ref{eq6}) becomes closer and
closer to the state in (\ref{eq5}).

In order to investigate the modifications on the quantum transition
rate induced by a sequence of $N$ interactions between $S$ and $M$
in a fixed time interval, let us consider $T$, divided as shown in
Fig 3.

\begin{figure}[h]
\centering
%\hspace{-3.5cm}
  \includegraphics[scale=0.65]{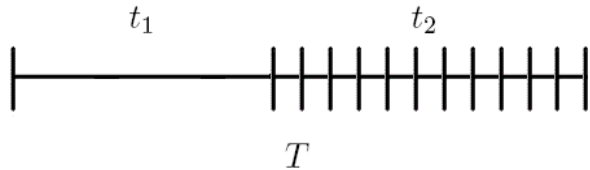}\\
  \caption{Graphic representation of the time interval $T$ division
  in $t_{1}$ when $S$ evolves freely and $t_{2}$ interval divided by $N$ interactions between $S$ and $M$.}
\end{figure}

In $t_{1}$ the system $S$ evolves freely, although, in $t_{2}$ a
sequence of $N$ interactions in performed.

If $N\rightarrow \infty$ the time interval between the interactions
tend to zero. Therefore, the dynamics becomes very similar to the
one in which $N$ consecutive interactions are performed in $t_{2}$.
The quantum transition rate after $N$ consecutive interactions in
$t_{2}$ can be written as:

\begin{equation}
\frac{dP_{1,0}^{(N)}}{dt}=-2\alpha(t_{1})\beta(t_{1})\cos^{N}(g).
\end{equation}

When $N\rightarrow \infty$, $\frac{dP_{1,0}^{(N)}}{dt}\rightarrow
0$. Thus, for a sequence composed of many measurements, the
incompleteness of each measurement is an irrelevant factor for the
inhibition of the quantum transition. The QZE independs of the
intensity of each correlations between $S$ and $M$, the quantum
transition rate becomes null in the limit $N\rightarrow \infty$,
provided that $|\cos(g)|\neq 1$.

To extend the analysis of the QZE for a finite sequence of
incomplete measurements, let us consider the graphics obtained by
numerical simulations on Fig 4. The curves show the probability
$P_{1,0}$ as function of $N$, for three values of $g$.

In a sequence of incomplete measurements, with $g$ between
$\frac{\pi}{2}<g<\pi$, two factors contribute to the inhibition of
the transition: derivative's signal inversion and decreasing of its
absolute value.

Let us carefully analise this two factors. At the initial instant
the system was prepared in $|1,0\rangle$, as we are investigating
the possibility of inhibition of the quantum transition
$|1,0\rangle\rightarrow |0,1\rangle$, in the time interval to be
considered, the probability $P_{1,0}$ is a decreasing function of
the time if no interactions between $S$ and $M$ are performed. After
the first measurement (with $\frac{\pi}{2}<g<\pi$), the curve
$P_{1,0} \times t$ inverts its behavior and becomes to increase (due
to the change on derivative signal), but with the absolute value of
transition rate reduced. After the second measurement, the curve
$P_{1,0} \times t$ decreases again and the absolute value of the
transition rate is even smaller. This effects continues as $N$
increases. The oscillation between increasing and decreasing
behavior of $P_{1,0} \times t$, as well as the successive reduction
on the absolute value of the transition rate, contribute to the
inhibition of the quantum transition.

In measurement sequences with $0<g<\frac{\pi}{2}$ only the reduction
on the absolute value of the transition rate contributes to the
inhibition of the transition. For this reason in Fig 4, the curve
with ($\cos(g)=-0,5$) shows the most rapid increase. When
($\cos(g)=0$) the measurement is complete (traditional QZE) the
transition rate becomes null after each measurement. For
($\cos(g)=-0,5$) the interactions induce only the reduction on the
absolute value of the transition rate. In the limit
$N\rightarrow\infty$ the three curves show the same behavior.

\begin{figure}[h]
\centering
%\hspace{-3.5cm}
  \includegraphics[scale=0.65]{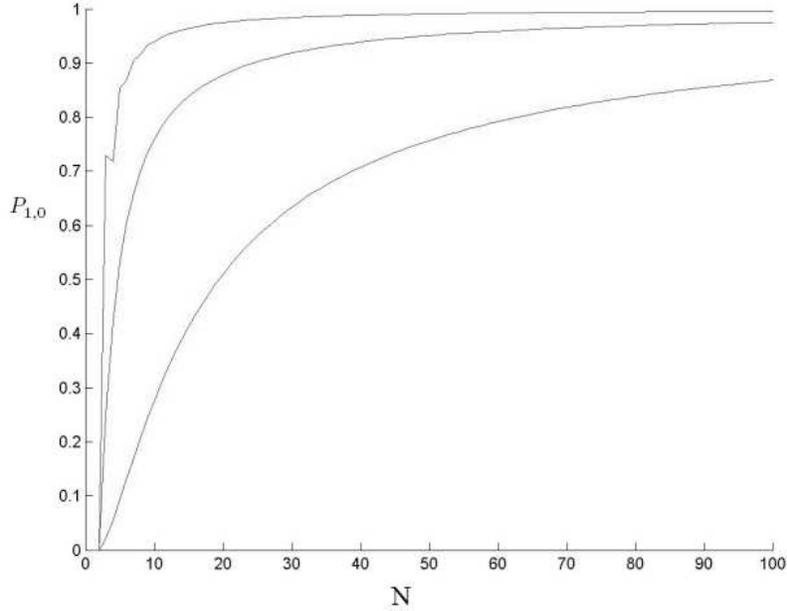}\\
  \caption{$P_{1,0}\times N$ for $g=\frac{3\pi}{4}$ (upper curve), $g=\frac{\pi}{2}$ (intermediated curve) and $g=\frac{\pi}{4}$ (bottom curve).}
\end{figure}

To summarize, in the present contribution, we have shown that the
QZE persistes even if the entanglement with the measuring probe is
not complete. We have also shown that when $\frac{\pi}{2}<g\leq\pi$
the interactions between $S$ and $M$ inhibit the transition in a
stronger sense than the traditional QZE.

M.C. Nemes and R. Rossi Jr. acknowledge financial support by CNPq.

\end{document}